\documentstyle[12pt,epsf]{article}
\newcommand{\beq}{\begin{equation}}
\newcommand{\eeq}{\end{equation}}
\newcommand{\vsin}{\vspace{-.3cm}}
\begin{document}
\pagestyle{myheadings}
\markright{S. Zapperi et. al., 
{\em Mat. Res. Soc. Proc. Vol.409, pag. 355 (1996)}}{

\begin{center}
{\bf MODELING ACOUSTIC EMISSION IN MICROFRACTURING PHENOMENA}
\end{center}
\noindent
S. ZAPPERI$^{\dag}$, A. VESPIGNANI$^{\ddag}$ AND 
H.E. STANLEY$^{\dag}$ 

\noindent
$^{\dag}$ Center for Polymer Studies and Department of Physics,
 Boston University, Boston, MA 02215 ,USA

\noindent 
$^{\ddag}$ Instituut-Lorentz, University of Leiden, P.O. Box 9506
The Netherlands.

\subsection*{\small ABSTRACT}
It has been recently observed that synthetic materials subjected
to an external elastic stress give rise to scaling phenomena in
the acoustic emission signal.  Motivated by this experimental
finding we develop a mesoscopic model in order to clarify the
nature of this phenomenon.
We model the synthetic material by an array
of resistors with random failure thresholds. The failure
of a resistor produces an decrease in the conductivity
and a redistribution of the disorder.
By increasing the applied voltage
the system organizes itself in a stationary state. The acoustic
emission signal is associated with the failure events. We find
scaling behavior in the amplitude of these events and in the
times between different events.
The model allows us to study the geometrical and topological
properties of the micro-fracturing process that drives the system
to the self-organized stationary state.

\subsection*{\small INTRODUCTION}

Acoustic Emission (AE)  is produced by sudden movements 
in stressed systems. Several experiments have  recently observed 
this phenomenon on very different length 
scales (from the largest scale of an 
earthquake to the smallest one of dislocation motions)
\cite{ppvac,dmp,ccc}.
Unfortunately, the AE analysis is a rather 
elicate technique since each external stress is unique and tests the 
whole sample. For instance, it is very difficult to obtain in this
way insight on the microscopic dynamics of the fracturing phenomena.
The statistical analysis, however, gives rise to the hypothesis 
that AE is generated by fracturing phenomena which 
are similar to critical points. Correlations 
develops leading to cascade events which drive the systems into a critical 
stationary state. For this reason, as  a working hypothesis, the  
mechanism of  Self-Organized Criticality (SOC) \cite{soc} has been invoked. 

The understanding of this statistical behavior calls for a model that 
can simulate the fracturing phenomenon. Unfortunately, the models usually 
considered describe the formation of a macroscopic crack \cite{hr}. 
These can therefore 
model AE of fracturing phenomena on mesoscopic scale culminating in a large 
event that change the system's properties dramatically \cite{th}. 
This is very different
from the stationary state generated by stressing the sample below the breaking
threshold of the system. In fact, in this case the fracturing phenomenon
produces an energy release that changes the physical properties in a 
non destructive way (like the passage to a different metastable state).

Here we show a statistical model for fracturing phenomena, where the 
rupture burst and the following energy release changes the properties 
but does not destroy the system. This allows us to obtain a stationary 
state for AE of which we can investigate the statistical properties in space,
time and magnitude.

\subsection*{\small THE MODEL}

The mesoscopic description of an elastic disordered medium
is obtained discretizing macroscopic elastic
equations. In the theory of linear elasticity,
these  equations relate the stress tensor $s_{\alpha\beta}$
to the strain tensor $\epsilon_{\gamma\delta}$ 
via the Hooke tensor $C_{\alpha\beta\gamma\delta}$. The full tensorial
formalism is quite heavy to handle numerically. A compromise
is obtained by considering scalar elastic equations. In fact the
phenomenology of fractures in scalar models captures many
essential features of more complex tensorial models.
Scalar elasticity is formally equivalent to electricity, 
provided one identifies the current $I$ with the stress, 
the voltage $V$ with the strain and the conductivity $\sigma$
with the Hooke tensor.

The discretization scheme we use corresponds to the study a
resistor network. For symmetry reason we will consider a
rotated square lattice. The disorder, due to the
inhomogeneity in the synthetic material, is introduced in the model
in the failure thresholds $I_c$ of the resistors. For simplicity we 
will use an uniform distribution.
The crucial part of the model is the breaking criterion, which
describe the dynamics of the micro-fracturing process. 
Typically  the breaking criterion is chosen so that if
the current flowing in a resistor exceed the failure threshold
the conductivity of the bond drops abruptly to zero. 
In this way the system develops a macroscopic crack 
and the lattice breaks apart. 
To describe the micro-fracturing phenomena in the stage
preceding the onset of the macroscopic crack it is
useful to consider the concept of damage. 
For a macroscopic elastic material, in which micro-fracturing
processes are taking place, 
the damage $D$ is a tensor relating the effective Hooke tensor $\hat{C}$
to the Hook tensor of the undamaged material. The damage is
defined as $D=I-C\hat{C}^{-1}$. For scalar elasticity the damage
is just a constant relating the effective resistance of
the damaged material to that of the undamaged one.
We generalize the concept of damage from the macroscopic
to the mesoscopic description, using it in the breaking criterion. 
When the current in a bond exceeds the threshold we impose a permanent
damage to the bond. In other words, the conductivity of
the bond drops by a factor $a=(1-D)$.

In the synthetic materials we are describing, after a micro-fracturing
event, local rearrangements take place. We model this effect by changing
at random the breaking threshold of the damaged bond as well as
those of the neighboring bonds. This rearrangement of the disorder
emphasize the probability of breaking successively neighboring
bonds. In crack models this process is enforced by imposing the connectivity 
of the crack or by similar rules.

\subsection*{\small SIMULATION RESULTS}

To simulate the model we start from an undamaged lattice where
the conductivities are equal to one for all the bonds. The
breaking threshold are chosen at random between zero and one.
We then impose an external voltage between two edges of the lattice
and we use periodic boundary conditions in the other direction.

The voltage is increased until the current in some bond 
exceed the threshold. When this happens we apply the breaking
rule and we check if subsequent failures occurs. In fact due
to the long range elastic interactions combined with the
redistribution of the disorder, a single failure can be
followed by other similar events, thus generating an avalanche.
We consider the number of bonds that break in an avalanche
to be proportional to the amplitude of the emitted acoustic signal.
In the early stage of the process only small avalanches occur
and the current carried by the material steadily increases.
In this stage the damage is scattered through the lattice in
a random uniform way. 
After some time when some damage has been accumulated into 
the system the activity starts to increase. Eventually the system
reaches a stationary state where the current does not
increase anymore. In other words the increase of the voltage 
is exactly balanced by the damage, in such a way that the current
is kept constant. In this state the damage is no longer homogeneously
scattered, but tends to be localized along lines.
We are modeling an ideal case in which a bond can suffer
a very big damage without breaking completely. In fact we can
slightly modify the model by introducing a threshold in the
conductivity after which the bond is consider broken
(i.e. the conductivity drops to zero). 
One can then easily understand that the regions of localized
damage are those where the macroscopic crack will eventually
form and this is indeed what we observe in our simulations.

The activity in this stationary state is highly fluctuating 
and the distribution of amplitude follows a power law. The
scaling region increases with the system size and the fact 
is a signature of an underlying critical state (see figure 1).

\begin{figure}[htb]
\centerline{
        \epsfxsize=16.0cm
        \epsfbox{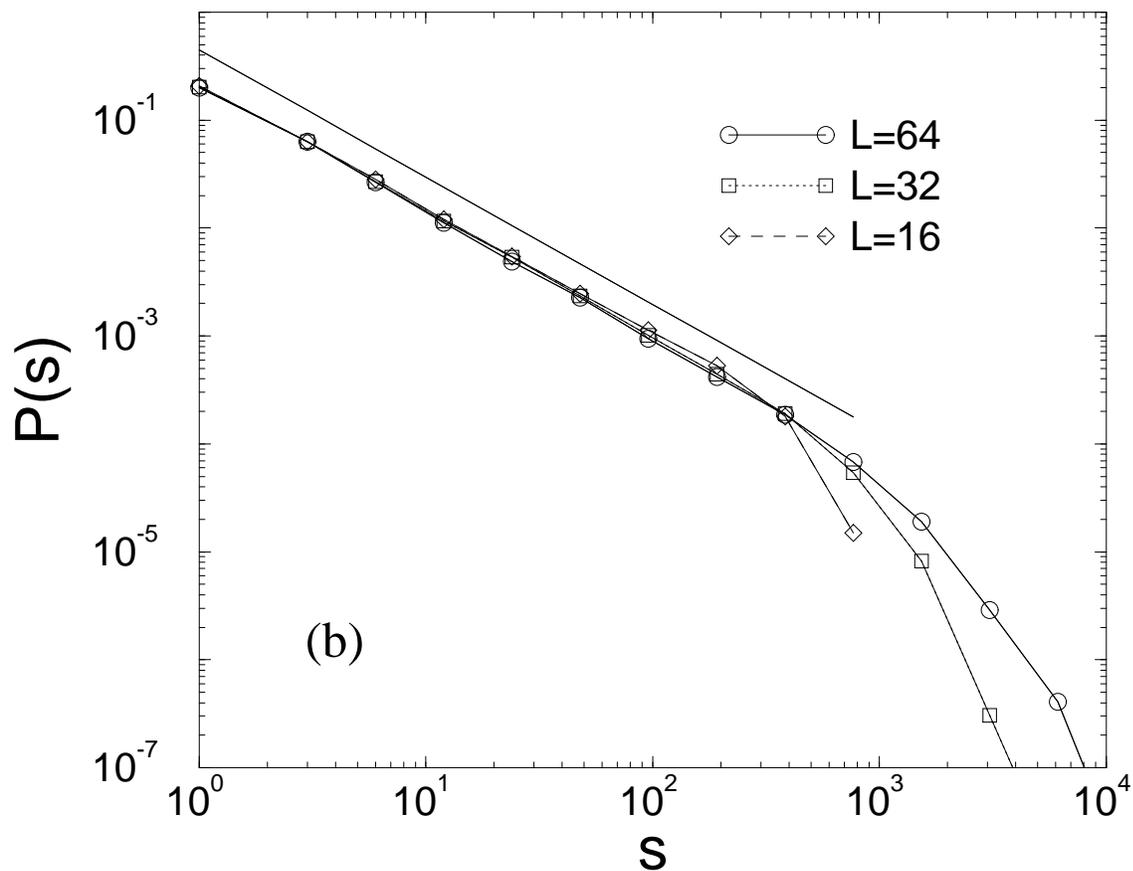}
        \vspace*{0.5cm}
        }
\caption{The avalanche distribution in the stationary state for
different system sizes.}
\label{fig:1}
\end{figure}

Another sign of the criticality of the system is provided 
by the distribution of time intervals between subsequent
avalanches. Since the voltage is increased linearly in time,
this corresponds to consider the distribution of voltage
increases $\Delta V$. The power law with slope close to $x=-1$
is reminiscent of the Omori \cite{omo} law for fore-shocks in earthquakes.

\subsection*{\small CONCLUSIONS}

In summary we have introduced a statistical model for 
fracturing phenomena. This model describes on a mesoscopic scale 
the energy release and the following rearrangements 
in the material produced by fracturing events. 
The simulations on the model show that the system organizes itself
in a stationary state where we can relate the AE signal with the 
rupture events. We investigate the statistical properties of 
rupture sequences during fracturing and their correlation properties.
We find that the stationary state develops critical correlations 
and scaling behavior through a self-organization process. 
This kind of analysis is usually considered in the study of 
experimentally detected AE signals and the proposed model could give 
important clues in the understanding of the general scale invariance 
of the fracturing phenomena. The Center for Polymer studies is
supported by NSF.
\vsin

\end{document}